\documentclass[prc,amsmath,twocolumn]{revtex4}

\usepackage{dcolumn}
\usepackage{graphicx}
\usepackage{color}

\newcommand*{\lrpart}{\tensor\partial}
\newcommand*{\s}[1]{/\llap{$#1$}}
\newcommand*{\eps}{\varepsilon}

\newcommand*{\kapp}{\kappa_p}
\newcommand*{\kapn}{\kappa_n}

\newcommand*{\kapr}{\kappa_\rho}

\newcommand*{\mr}{{m_\rho}}

\newcommand*{\gammu}{\gamma^\mu}
\newcommand*{\gamnu}{\gamma^\nu}
\newcommand*{\gamF}{\gamma^5}
\newcommand*{\munu}{{\mu\nu}}
\newcommand*{\K}{\ensuremath{\mathsf K}}
\newcommand*{\tf}{\ensuremath{\tilde f}}
\newcommand*{\dop}[2]{#1\cdot#2}
\newcommand*{\GeV}{\ensuremath{\;\mbox{GeV}}}

\newcommand*{\La}{\ensuremath{\mathcal L}}

\newcolumntype{x}[1]{D..{#1}}
\newcolumntype{C}{>{$}c<{$}}
\newcolumntype{R}{>{$}r<{$}}

\newcommand*{\tblref}[1]{Table~\ref{tbl:#1}}
\newcommand*{\tbllab}[1]{\label{tbl:#1}}
\renewcommand*{\eqref}[1]{Eq.~(\ref{eq:#1})}
\newcommand*{\eqlab}[1]{\label{eq:#1}}
\newcommand*{\figref}[1]{Fig.~\ref{fig:#1}}
\newcommand*{\figlab}[1]{\label{fig:#1}}
\newcommand*{\secref}[1]{Section~\ref{sec:#1}}
\newcommand*{\seclab}[1]{\label{sec:#1}}

\def\VYP#1#2#3{{\bf #1}, #3 (#2)}  

\def\PRC#1#2#3{Phys.~Rev.~C~\VYP{#1}{#2}{#3}}

\begin{document}

\title{Channel coupling effects in $\rho$-meson photoproduction.}

\author{A. Usov}
\email{usov@kvi.nl}
\author{O. Scholten}
\email{scholten@kvi.nl}
\affiliation{Kernfysisch Versneller Instituut, University of Groningen,
9747 AA, Groningen, The Netherlands}

\begin{abstract}
  We investigate $\rho$-meson photoproduction in a coupled-channels
  formulation. It is shown that channel coupling effects are large and
  account for discrepancies observed in several single-channel
  treatments.
\end{abstract}
\maketitle

\section{Introduction}

A number of experiments on vector meson production at low energies
\cite{Anci00,Luka01,Batt01,Batt03} performed in recent years have
greatly stimulated theoretical research on the subject~\cite{Pen02a,
  Pen02b,Shkl05,Zhao98b,Zhao00,Zhao01,Titov02,Titov03,Oh02,Oh04}. The
investigation of vector meson production is expected to provide
insight in the problem of ``missing resonances''~\cite{Caps00}, as
many of the predicted resonant states are believed to couple weakly to
the pion channel, but should be visible in other reaction channels.
Furthermore, in our previous analysis of photo-induced kaon
production~\cite{Usov05} we saw that the inclusion of the pion-induced
$\rho$-meson production channel has a strong influence on pion
scattering and $K-\Lambda$ photoproduction cross-sections, but the
detailed analysis of $\rho$-meson production was not done at that
time.  Therefore, in the current paper, we concentrate primarily on
the analysis of $\rho$-meson production and it's effects in a coupled
channels calculation.

Most of the work done on vector meson production concentrated
primarily on $\omega$-meson production. While most of the analyses
performed so far employed tree-level
models~\cite{Zhao98b,Zhao00,Zhao01,Titov02,Titov03,Oh04}, there are
strong indications for the importance of channel coupling effects
(loop corrections) for $\omega$-meson production~\cite{Oh02}. A number
of calculations performed using the $\K$-matrix
approach~\cite{Pen02a,Pen02b,Shkl05} included the $\omega$-production
channels, but the magnitude of the contributions due to channel
coupling was not investigated explicitly. The analyses of $\omega$-
and $\rho$-meson production channels in the ``effective quark model
Lagrangian'' approach~\cite{Zhao98b} have found a good simultaneous
description of the data at the time, however these do not include the
coupling to the pion sector, which was found to be
essential~\cite{Oh02}. In general, tree-level models have
difficulties~\cite{Oh04} describing the differential cross-section in
the region of high momentum transfer. We show that coupled-channels
treatment of the photo-induced $\rho$-meson production is able to
account for a major part of the discrepancies observed in a
single-channel model. The coupling to the pion-nucleon state is by far
the most important ingredient.

In \secref{model} we give a brief outline of our model and in
\secref{results} we present and discuss the results of the
calculation.

\section{Model}\seclab{model}

The model used in this work is based on an effective Lagrangian
formalism. The detailed description of the model can be found in our
previous paper~\cite{Usov05}, here we will summarize it, give a
description of the $\gamma+N \to \rho+N$ and $\pi+N \to \rho+N$
reaction channels and specify the modifications made to other reaction
channels in order to extend the applicability of the model to higher
energies.

Our model is based on the \K-matrix formalism to implement channel
coupling. A discussion of the \K-matrix formalism can be found in
ref.~\cite{Usov05} and references therein. The use of the \K-matrix
formalism allows to generate an infinite, non-perturbative set of loop
corrections for states explicitly included in the model, while obeying
a number of symmetries like gauge invariance, unitarity and crossing
symmetry. The model space used in the present investigation is formed
by $K-\Lambda$, $K-\Sigma$, $\phi-N$, $\eta-N$, $\gamma-N$, $\pi-N$
and $\rho-N$ states. A number of non-strange resonances are included
in s- and u-channel contributions. For the complete list of the
included resonances and their parameters we refer to our previous
publication~\cite{Usov05}.

Parameters entering the calculation are mostly unchanged as compared
to what was presented in our previous calculation~\cite{Usov05}. The
most important exceptions are the $NN\omega$ coupling constant and a
few cut-offs in the pion photoproduction channel. In
\tblref{parameters} we quote the values of the parameters relevant for
the current discussion.

In the following sections we use a notation where $p$, $k$, $p'$ and
$-q$ correspond to the initial proton, $\rho$-meson and final proton
and photon(pion) 4-momenta respectively. Indices $\mu$ and $\nu$ label
the photon- and $\rho$-meson polarization indices. We assume that the
meson momenta are directed into the vertex, so that energy-momenta
conservation reads as $p+k=p'-q$.

The complete Lagrangian is given in ref.~\cite{Usov05} and here
we quote only the terms relevant for the current discussion.

\newcommand*{\YYv}[1]
    {\frac{(\chi #1 + i \s\partial #1 /2M) \cdot \vec\tau}
         {\chi+1} \gamF}
\newcommand*{\YY}[1]
    {\frac{\chi #1 + i \s\partial #1 /2M}
         {\chi+1} \gamF}
\newcommand*{\XXv}[1]
    {\left( \gamma_\mu \vec {#1}^\mu 
      +\frac{\kappa_{#1}}{2M} \sigma_{\mu\nu} \partial^\nu \vec {#1}^\mu
      \right) \cdot \vec\tau}
\newcommand*{\XX}[1]
    {\left( \gamma_\mu {#1}^\mu 
      +\frac{\kappa_{#1}}{2M} \sigma_{\mu\nu} \partial^\nu {#1}^\mu
      \right)}
\newcommand*{\EPS}[4]
    {\left(\eps_{\mu\nu\rho\sigma}
      (#1^\rho #2^\mu) (#3^\sigma #4^\nu) \right)}

\begin{equation}\eqlab{Lagrangian}
  \begin{aligned}
    \La_{NN\pi} &= i g_{NN\pi} \bar N \YYv{\vec\pi} N \\
    \La_{NN\eta} &= i g_{NN\eta} \bar N \YY{\eta} N \\
    \La_{NN\sigma} &= -g_{NN\sigma} \bar N \sigma N \\
    \La_{NN\rho} &= -g_{NN\rho} \bar N \XXv{\rho\,} N \\
    \La_{NN\omega} &= -g_{NN\omega} \bar N \XX{\omega} N \\
    \La_{NN\gamma} &= -e \bar N 
       \left( \frac{1+\tau_0}{2} \gamma_\mu A^\mu 
         +\frac{\kappa_\tau}{2M} \sigma_{\mu\nu} \partial^\nu A^\mu
       \right) N \\
    \La_{\rho\pi\pi} &= - g_{\rho\pi\pi}
             \vec\rho_\mu \cdot (\vec\pi \times \lrpart^\mu \vec\pi) /2 \\
    \La_{\gamma\pi\pi} &= e \eps_{3ij} A_\mu
       (\pi_i \lrpart^\mu \pi_j) \\
    \La_{\rho\gamma\pi} &= e \frac{g_{\rho\gamma\pi}}{m_\pi} \dop{\vec\pi}{
        \EPS{\partial}{A}{\partial}{\vec\rho\,}} \\
    \La_{\omega\gamma\pi} &= e \frac{g_{\omega\gamma\pi}}{m_\pi} \pi^0
        \EPS{\partial}{A}{\partial}{\omega} \\
    \La_{\rho\gamma\eta} &= e \frac{g_{\rho\gamma\eta}}{m_\pi} \eta
        \EPS{\partial}{A}{\partial}{{\rho^0}} \\
    \La_{\rho\gamma\sigma} &= e \frac{g_{\rho\gamma\sigma}}{m_\rho}
        (\partial^\mu \rho^\nu \partial_\mu A_\nu
          - \partial^\mu \rho^\nu \partial_\nu A_\mu ) \\
    \La_{\rho\rho\gamma} &= 2e \big(
        A^\mu (\partial_\mu \rho_\nu) \tau_0 \rho^\nu
        -(\partial^\nu A^\mu) \rho_\nu \tau_0 \rho_\mu \\
        &\quad +(\partial^\nu A^\mu) \rho_\mu \tau_0 \rho_\nu \big)
  \end{aligned}
\end{equation}

For all channels discussed here we include a complete set of s-, u-
and t-channel Born diagrams. In the photon- and $\rho$-nucleon
vertices form-factors for the magnetic moment contributions are
included. The form-factors for the convection current contributions
are implemented through the use of counter terms.

Currently we do not include direct matrix elements between the
$\rho-N$ and kaon sector. Because of the relatively weak coupling, as
compared to the pion sector, this will not affect our main results.
Also the final state interaction for the $\rho-N$ state is not
included. In the photo-induced $\eta$ production channel we have taken
into account an additional $\rho$-meson t-channel contribution in
order to keep consistency with other channels.  The value of the
coupling constant $g_{\rho\eta\gamma}$ was taken from decay data.

\begin{table}
  \caption{\tbllab{parameters} Parameters summary table}
  \begin{ruledtabular}
  \begin{tabular}{C|C|C|C}
    g_{NN\pi}      & 13.47 & g_{\rho\pi\pi}      & 6 \\
    g_{NN\eta}     & 3     & g_{\rho\pi^0\gamma} & -0.12 \\
    g_{NN\sigma}   & 10    & g_{\rho\pi^\pm\gamma}&-0.10 \\
    g_{NN\rho}     & 2.2   & g_{\rho\eta\gamma}  & 0.22 \\
    g_{NN\omega}   & 3.0   & g_{\omega\pi\gamma} & 0.32 \\
                   &       & g_{\rho\sigma\gamma}& 12 \\
  \end{tabular}
  \end{ruledtabular}
\end{table}


\subsection{The $\gamma+N \to \rho+N$ reaction channel}\seclab{gam-rho}
\subsubsection{Born-level contributions and contact terms}

Since we do not assign form factors to the convection current part of
the $\rho$ and photon vertices at the Born level, the pure convection
current contribution is gauge invariant. However, the terms where one
of the vertices is magnetic, are suppressed by form-factors and are
subject to a gauge restoration procedure. The contact term required to
restore gauge invariance reads
\begin{multline}\eqlab{gam-rho-ct-rho-0}
  C_{\rho^0}^\munu =
   -e \frac{g_{NN\rho}\kapr}{2m} \sigma^{\nu\lambda} k_\lambda
      \Big[ (2p+q)^\mu F^M_m(s)\tf^M_m(u) \\
        + (2p'-q)^\mu F^M_m(u)\tf^M_m(s) \Big] \\
   -\frac{g_{NN\rho}\kapp}{2m} \sigma^{\mu\lambda} q_\lambda
      \Big[ (2p'-k)^\nu F^M_m(s)\tf^M_m(u) \\
        + (2p+k)^\nu F^M_m(u)\tf^M_m(s) \Big],
\end{multline}
\begin{multline}\eqlab{gam-rho-ct-rho-p}
  C_{\rho^+}^\munu =
   -e \frac{g_{NN\rho}\kapr}{2m} \sigma^{\nu\lambda} k_\lambda
      \Big[ (2k-q)^\mu F^M_m(s)\tf^M_\mr(t) \\
        + (2p'-q)^\mu F^M_\mr(t)\tf^M_m(s) \Big] 
   -e \frac{g_{NN\rho}\kapr}{2m} \sigma^\munu F^M_\mr(t) \\
   -g_{NN\rho}\frac{\kapr F^M_\mr(t) - \kapp F^M_m(s) + \kapn F^M_m(u)}{2m}
      \sigma^{\mu\lambda} q_\lambda \frac{k^\nu}{\mr^2}.
\end{multline}
Where $\tf^M_m(s)$ is defined as $\tf^M_m(s) = (1-F^M_m(s))/(s-m^2)$
and the form-factors are given in \eqref{gam-rho-ff-magn}. Note, that
the last term of \eqref{gam-rho-ct-rho-p} does not contribute to
physical observables and serves only gauge invariance restoration
purposes. We should note that this contact term is not unique and the
exact functional form strongly depends on the ordering of terms in the
minimal substitution procedure~\cite{Usov05,Kondr00,Kondr00b}. The
present functional form is chosen to resemble the structure of the
contact term used in kaon production, see Eq.~(11a) and Eq.~(11b) of
ref.~\cite{Usov05}. To present this contact term in a more readable
way, we have given it for the $\rho^0$ and $\rho^+$ production
channels separately.

In order to reproduce experimental data at high energies one also has
to suppress the convection current contributions with form-factors.
For this we have chosen to implement a prescription similar to the one
proposed by Davidson and Workman~\cite{Davi01}. Written as a
gauge-invariant contact term~\cite{Usov05} it reads
\begin{multline}\eqlab{gam-rho-ct-DW}
  C_{DW}^\munu =
    - e g_{NN\rho} \Big[
     4\gamnu \big(\dop{p}{q} p'^\mu - \dop{p'}{q} p^\mu \big) \\
     -\big(2\dop{p}{q} (\gammu \s q \gamnu) 
       + 2\dop{p'}{q} (\gammu \s q \gamnu)\big)
    \Big] \\ \times
    \left((1-e_\rho) \tf_m(s)\tf_m(u) 
      - e_\rho \tf_m(s) \tf_\mr(t) \right),
\end{multline}
where $\tf_m(s) = (1-F_m(s))/(s-m^2)$. This also gave good results
for kaon production.

In order to achieve a better suppression of the u-channel
contributions in the region of high momentum transfer we have chosen
to implement ``modified'' form-factors, as was done in the case of kaon
production.  The s-channel form-factor is redefined primarily for
consistency considerations. This has hardly any effect on the results
of the calculation, which are very similar to those obtained with the
unmodified s-channel form-factor.
\begin{equation}\eqlab{gam-rho-ff}
  \begin{gathered}
    F_m(s) = \frac{s \Lambda^2}{\big(\Lambda^2 + (s-m^2)^2 \big) m^2}, \\
    F_m(u) = \frac{u \Lambda^2}{\big(\Lambda^2 + (u-m^2)^2 \big) m^2}, \\
    F_m(t) = \frac{  \Lambda^2}{\Lambda^2 + (t-m^2)^2}. 
  \end{gathered}
\end{equation}
The magnetic moments are suppressed by the quadrupole form-factors
defined as
\begin{equation}\eqlab{gam-rho-ff-magn}
  \begin{gathered}
    F^M_m(s) = \Bigg(\frac{s \Lambda^2}{\big(\Lambda^2 + (s-m^2)^2 \big) m^2}\Bigg)^2, \\
    F^M_m(u) = \Bigg(\frac{u \Lambda^2}{\big(\Lambda^2 + (u-m^2)^2 \big) m^2}\Bigg)^2, \\
    F^M_m(t) = \Bigg(\frac{  \Lambda^2}{\Lambda^2 + (t-m^2)^2}\Bigg)^2. 
  \end{gathered}
\end{equation}
All form-factors entering
equations~(\ref{eq:gam-rho-ct-rho-0}-\ref{eq:gam-rho-ct-DW}) use the
same, relatively hard, cut-off of $\Lambda=0.8 \GeV^2$.

\subsubsection{Additional t-channel contributions}

In addition to the Born-level contributions described above we include
a number of t-channel contributions. At the higher energies, t-channel
Regge exchanges are becoming more important and thus we include the
Regge propagator for the dominant $\sigma$-meson t-channel exchange.
In addition $\pi$ and $\eta$ t-channel exchanges are included, however
these are of less importance.

Matrix elements for $\eta$ and pion-exchange diagrams, following from
the interaction Lagrangian (\eqref{Lagrangian}), are given by
\begin{multline}
  M_\phi^\munu =
    i g_{NN\phi} \frac{e g_{\rho\phi\gamma}}{m_\pi}
      \eps^{\munu\alpha\beta} k_\alpha q_\beta
      \frac{1}{t-M_\phi^2} \\ \times
      \bar u(p') \gamF \frac{\s p' - \s p}{2M_N} u(p)
      F_{m_\phi}(t),
\end{multline}
where $\phi$ is either $\eta$ or $\pi$ and $F_{m_\phi}(t)$ is a
dipole-type form-factor as is presented in~\eqref{gam-rho-ff}. For
both $\eta$ and $\pi$ contributions a hard cut-off of $0.8 \GeV^2$
was used (the exact value does not matter much, and was chosen for
consistency reasons).

The matrix element corresponding to $\sigma$-meson exchange,
following from the interaction Lagrangian reads
\begin{multline}
  M_\sigma^\munu =
    g_{NN\sigma} \frac{g_{\rho\sigma\gamma}}{M_\rho}
      (\dop{k}{q} g^\munu - q^\nu k^\mu)
      P_\sigma(s,t) \\ \times
      \bar u(p') u(p) F_{m_\sigma}(t),
\end{multline}
where $F_{m_\sigma}(t)$ is the same dipole t-channel form-factor
of~\eqref{gam-rho-ff} with cut-off set to $1 \GeV^2$. The Regge
propagator $P_\sigma(s,t)$ is defined as~\cite{Cano03,Guid97}
\begin{equation}
  P_\sigma(s,t) = \left(\frac{s}{s_0}\right)^{\alpha_\sigma(t)}
     \frac{\pi \alpha'_\sigma}{\Gamma(1+\alpha_\sigma(t))}
     \frac{e^{-i\pi \alpha_\sigma(t)}}{\sin(\pi \alpha_\sigma(t))},
\end{equation}
where $\alpha_\sigma(t)$ is $\sigma$-meson Regge trajectory defined as
$\alpha_\sigma(t) = \alpha^0_\sigma + \alpha'_\sigma t$ with a slope
of $\alpha'_\sigma = 0.7 \GeV^{-2}$ and intercept of $\alpha^0_\sigma
= -\alpha'_\sigma m_\sigma^2 = -0.4$. The coupling constants
$g_{NN\sigma}$ and $g_{\rho\sigma\gamma}$ are given
in~\tblref{parameters}. The reference scale $s_0$ is conventionally
chosen to be $1 \GeV^2$.

\begin{figure*}
  \includegraphics[angle=90,width=1.0\textwidth]{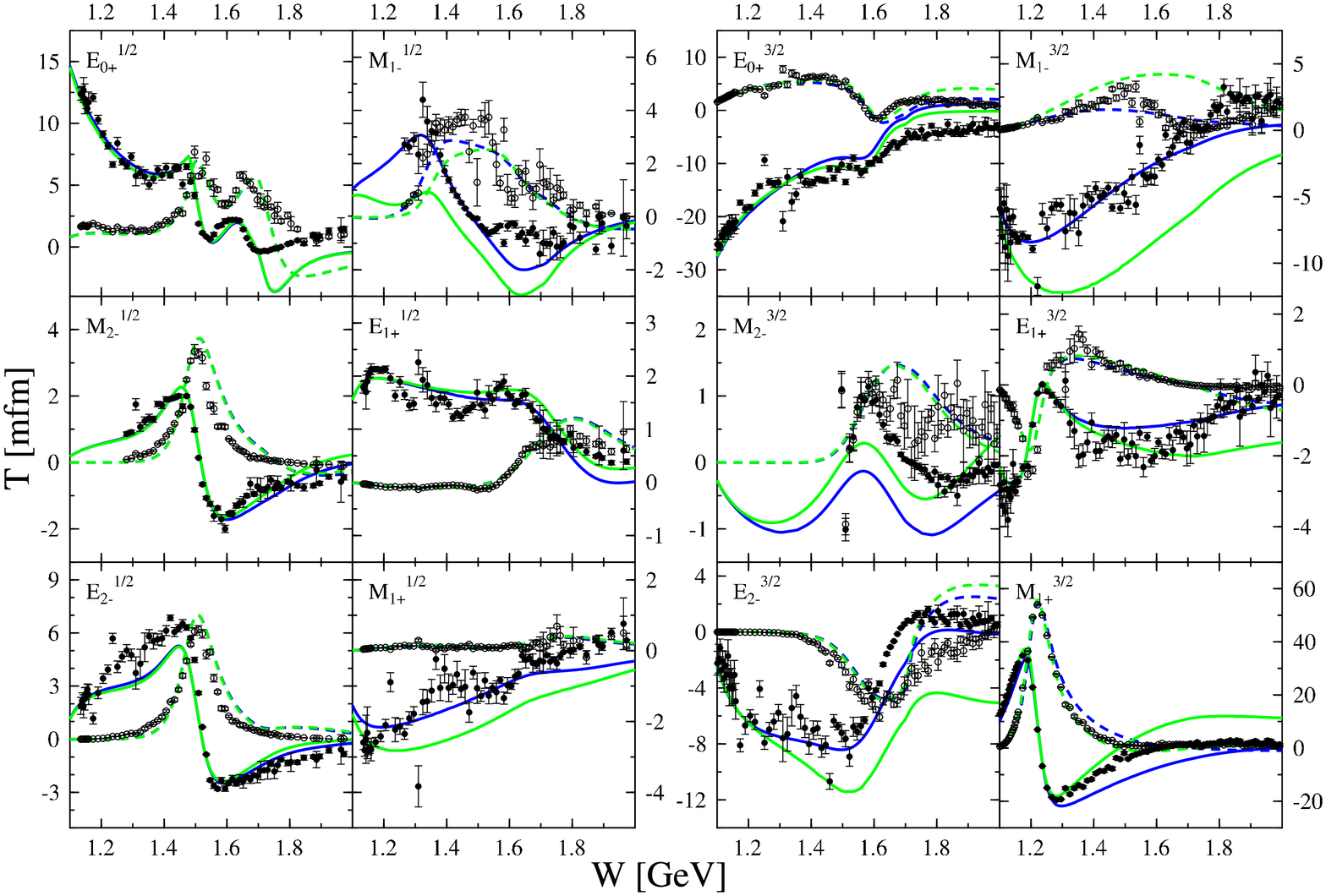}
  \caption{[Color online] Partial wave decomposition of the 
    $\gamma+N \to \pi+N$ reaction channel. Solid and dashed lines
    correspond to the real and imaginary parts of the matrix element
    respectively. Darker (blue) lines shows the results of the new
    calculation, while lighter (green) lines correspond to the
    calculation where the tree-level $\gamma+N \to \pi+N$ reaction
    channel agrees with that of \cite{Usov05}. The data are taken from
    the on-line database of the VPI group~\protect\cite{SAID}.}
  \figlab{gam-pi-pw}
\end{figure*}

\subsection{The $\pi+N \to \rho+N$ reaction channel}

\newcommand*{\Smatr}{(1-\dop{\vec\tau_N}{\vec\tau_\pi})}
\newcommand*{\Umatr}{(1+\dop{\vec\tau_N}{\vec\tau_\pi})}
\newcommand*{\Tmatr}{(\dop{\vec\tau_N}{\vec\tau_\pi})}
\newcommand*{\Cmatr}{\frac{5\dop{\vec\tau_N}{\vec\tau_\pi}-2}{6}}

Similar to the $\gamma+N \to \rho+N$ reaction channel, the Born-level
contributions include form-factors for the magnetic moment while the
convection current contributions are suppressed by counter terms.

The $\rho$-meson, unlike the photon, couples with a different strength to
the nucleon (in s- and u-channel contributions) and pion (t-channel
contribution). As a result gauge invariance is obeyed only when
$g_{NN\rho}=2g_{\rho\pi\pi}$. To ensure gauge invariance for arbitrary
values of the coupling constants we have introduced following contact
term (putting $\chi=0$)
\begin{multline}\eqlab{pi-rho-ct}
  C^\nu = \frac{1}{4m} \Big(
     -g_{NN\pi} (2g_{NN\rho} - g_{\rho\pi\pi})
       (\s q + \s k)\gamF \\
     +2g_{NN\rho} g_{NN\pi}
       \s k \gamF
     \Big) \frac{k^\nu}{m_\rho^2} \Tmatr.
\end{multline}
This particular choice for the contact term gives a vanishing
contribution to physical observables. However the choice of the
contact term is ambiguous, where the ambiguity can be expressed as
\begin{equation}\eqlab{pi-rho-ct2}
  C^\nu = g \frac{\gamma^\nu - \s k k^\nu/m_\rho^2}{2m} \gamF
    (a+b\dop{\vec\tau_N}{\vec\tau_\pi}),
\end{equation}
where $g$, $a$ and $b$ are arbitrary parameters. To fix the parameters
we use the fact that a similar contact term appeared in kaon
photoproduction. Exploiting the analogy between $\rho$-meson
production and photo-induced kaon production we have fixed the
parameters for our full calculation as $g=g_{NN\pi} g_{\rho\pi\pi}$,
$a=0$ and $b=1/2$. Later in \secref{results} we discuss the effects of
this contact term on the results in more detail.

To implement form-factors for the convection current contributions we
rewrite them first in terms of gauge-invariant amplitudes (these do
not exactly correspond to ${\mathsf M}_i$ of photo-induced kaon
production, although they are quite similar). For the part of the
convection current contribution which corresponds to the ${\mathsf
  A}_2$ amplitude of kaon production, we implement the DW-style
suppression scheme~\cite{Davi01}. However, due to the more complicated
structure (in the general case of $g_{NN\rho} \ne 2g_{\rho\pi\pi}$) we
assign them a common overall form-factor $F^{DW}_{sut}$~\cite{Davi01}.
The corresponding contact term reads
\begin{multline}
  C_{DW}^\nu= \Bigg[
     -\frac{g_{NN\rho} g_{NN\pi}}{4m}
       \s q \gamF \Bigg(
       \frac{(2p+k)^\nu}{s-m^2} \Smatr \\
       +\frac{(2p'-k)^\nu}{u-m^2} \Umatr
       -4\frac{k^\nu}{m_\rho^2} \Tmatr
       \Bigg) \\
     -\frac{g_{NN\pi} g_{\rho\pi\pi}}{4m}
       (\s q + \s k)\gamF
       \frac{2q^\nu-2\dop{q}{k}k^\nu/m_\rho^2}{t-m^2}
       \Tmatr \Bigg] \\ \times
       \tf_m(s) \tf_m(u) \tf_{m_\pi}(t)
                     (s-m^2) (u-m^2) (t-m_\pi^2).
\end{multline}

The remaining contributions of the convection current are suppressed
by individual form-factors through the inclusion of the following
contact term
\begin{multline}
  C^\nu=
     -\frac{g_{NN\rho} g_{NN\pi}}{4m} \Bigg[
       \s q\gamF \frac{\s k \gamma^\nu - \gamma^\nu \s k}{2} \tf_m(s) \Smatr \\
       +\frac{\s k \gamma^\nu - \gamma^\nu \s k}{2} \s q\gamF \tf_m(u) \Umatr
       \Bigg].
\end{multline}

In addition, for the reasons discussed in the results section, we
include an extra contact term of the following form
\begin{equation}\eqlab{pi-rho-ct3}
  C^\nu = -g_{NN\pi} g_{\rho\pi\pi} \gamF
     \frac{2q^\nu-2\dop{q}{k}k^\nu/m_\rho^2}{m_\rho^2}
     (a+b\dop{\vec\tau_N}{\vec\tau_\pi}) F_m(t),
\end{equation}
where $a=1/3$ and $b=5/6$.

In the $\pi+N \to \rho+N$ reaction channel we do not use modified
u-channel form-factors. The reason is that the effect of their
inclusion is quite small and we opted for the simpler choice of dipole
form-factors.

\subsection{The $\gamma+N \to \pi+N$ reaction channel}

To improve the cross-sections at high energy we have introduced
additional quadrupole form-factors for the magnetic moment
contributions and used a ``modified'' u-channel form-factor. In
addition, the cut-off value for the form-factors entering ``core''
contributions (s-, u-, t-channel contributions, gauge-invariance
restoration and DW-contact terms) was reduced to $0.8 \GeV^2$ and the
$NN\omega$ coupling constant was reduced by a substantial factor, as
compared with the previous calculation.

\begin{figure}
  \includegraphics[angle=90,width=1.0\linewidth]{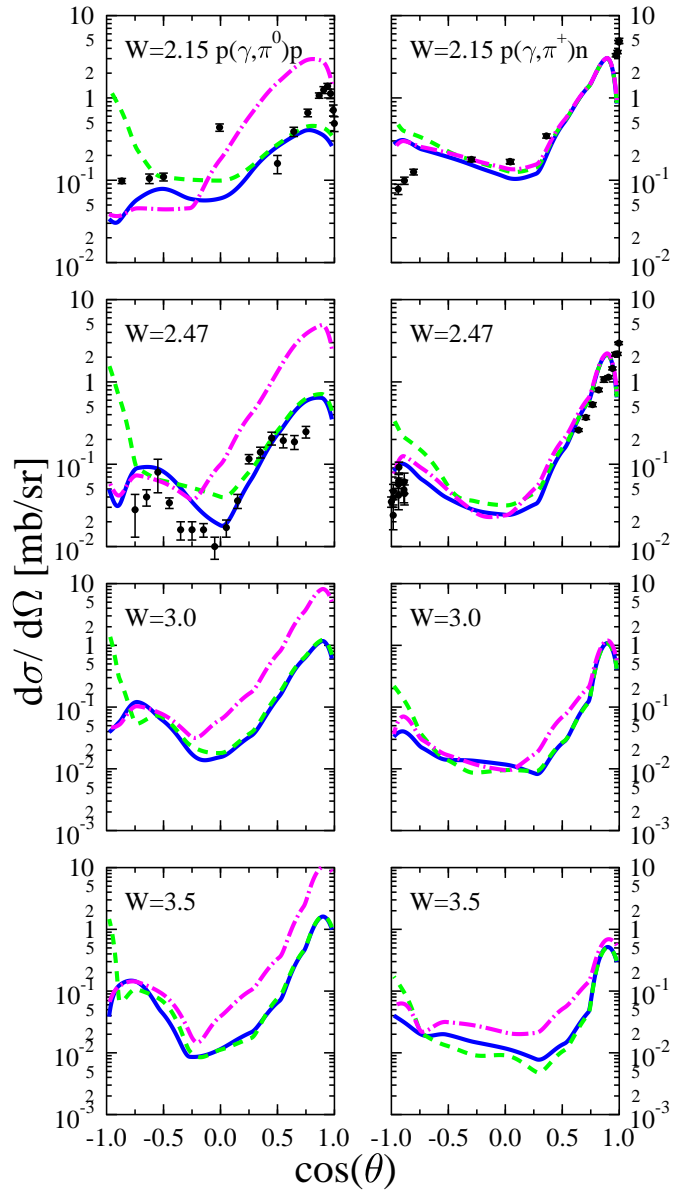}
  \caption{[Color online] Differential cross-sections of the
    $\gamma+N \to \pi+N$ reaction channel. See the text for the
    explanation of different curves. The data are taken from the
    on-line database of the VPI group~\protect\cite{SAID}.}
  \figlab{gam-pi-diff}
\end{figure}

\begin{figure*}
  \includegraphics[angle=90,width=1.0\textwidth]{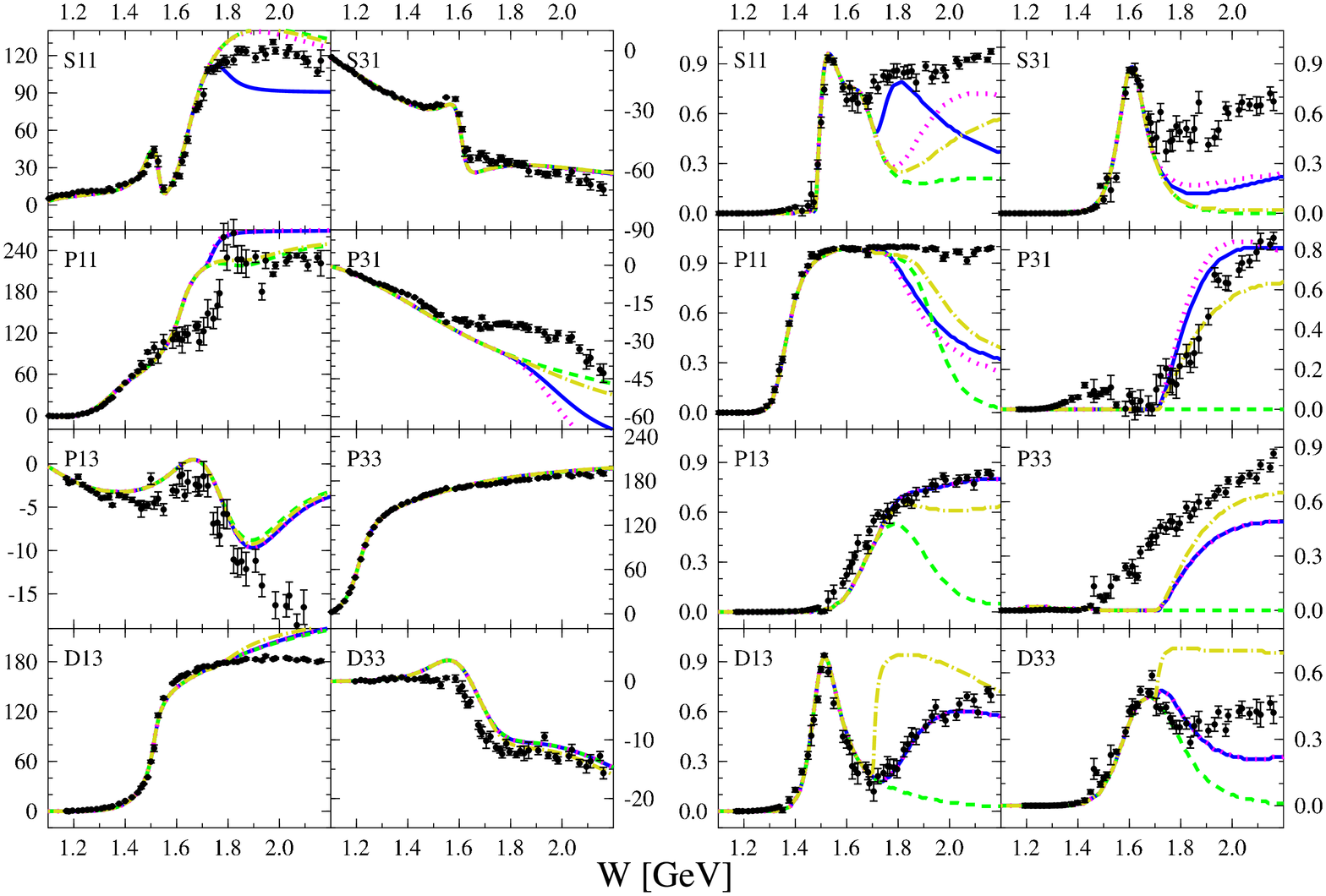}
  \caption{[Color online] Partial wave decomposition for the
    $\pi+N \to \pi+N$ reaction channel. The left plane shows
    per-partial wave phase shifts, and inelasticities are shown in the
    right plane. The solid (blue) line corresponds to the full
    calculation. The dashed (green) line shows the calculation where
    all reaction channels including $\rho-N$ final state were
    excluded.  The dotted (magenta) line denotes the calculation where
    an additional contact term of~\eqref{pi-rho-ct2} is excluded from
    the $\pi+N \to \rho+N$ reaction channel (note, that in spin-$3/2$
    partial waves it agrees with the full calculation). The
    dash-dotted (yellow) line corresponds to the calculation where the
    contact term of~\eqref{pi-rho-ct3} is excluded in the $\pi+N \to
    \rho+N$ reaction channel. The data are taken from the on-line
    database of the VPI group\protect\cite{SAID}.}  \figlab{pi-pi-pw}
\end{figure*}

\begin{figure}
  \includegraphics[angle=0,width=1.0\linewidth]{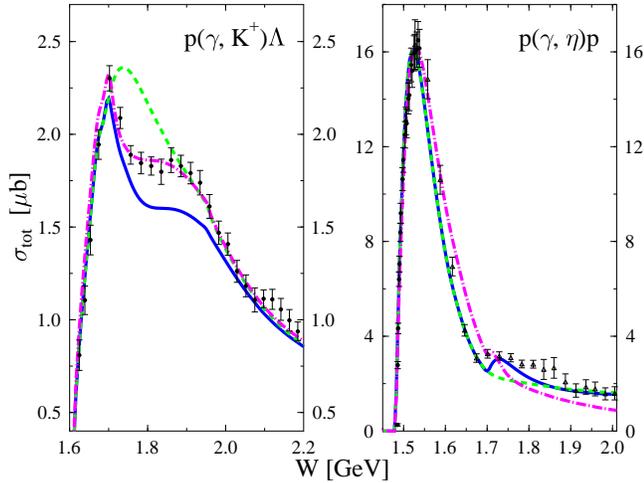}
  \caption{[Color online] The effect of the contact term
    of~\eqref{pi-rho-ct2} on kaon and $\eta$ photoproduction. Solid
    (blue) line shows full calculation. Dashed (green) line
    corresponds to the calculation where contact term
    of~\eqref{pi-rho-ct2} is excluded in $\pi+N \to \rho+N$ reaction
    channel. The dash-dotted (magenta) line shows the results of the
    calculation as presented in our previous paper~\cite{Usov05}. The
    data are taken from Refs.~\cite{Glan04,Cred05}.} \figlab{gam-kl-tot}
\end{figure}

\begin{figure}
  \includegraphics[angle=90,width=1.0\linewidth]{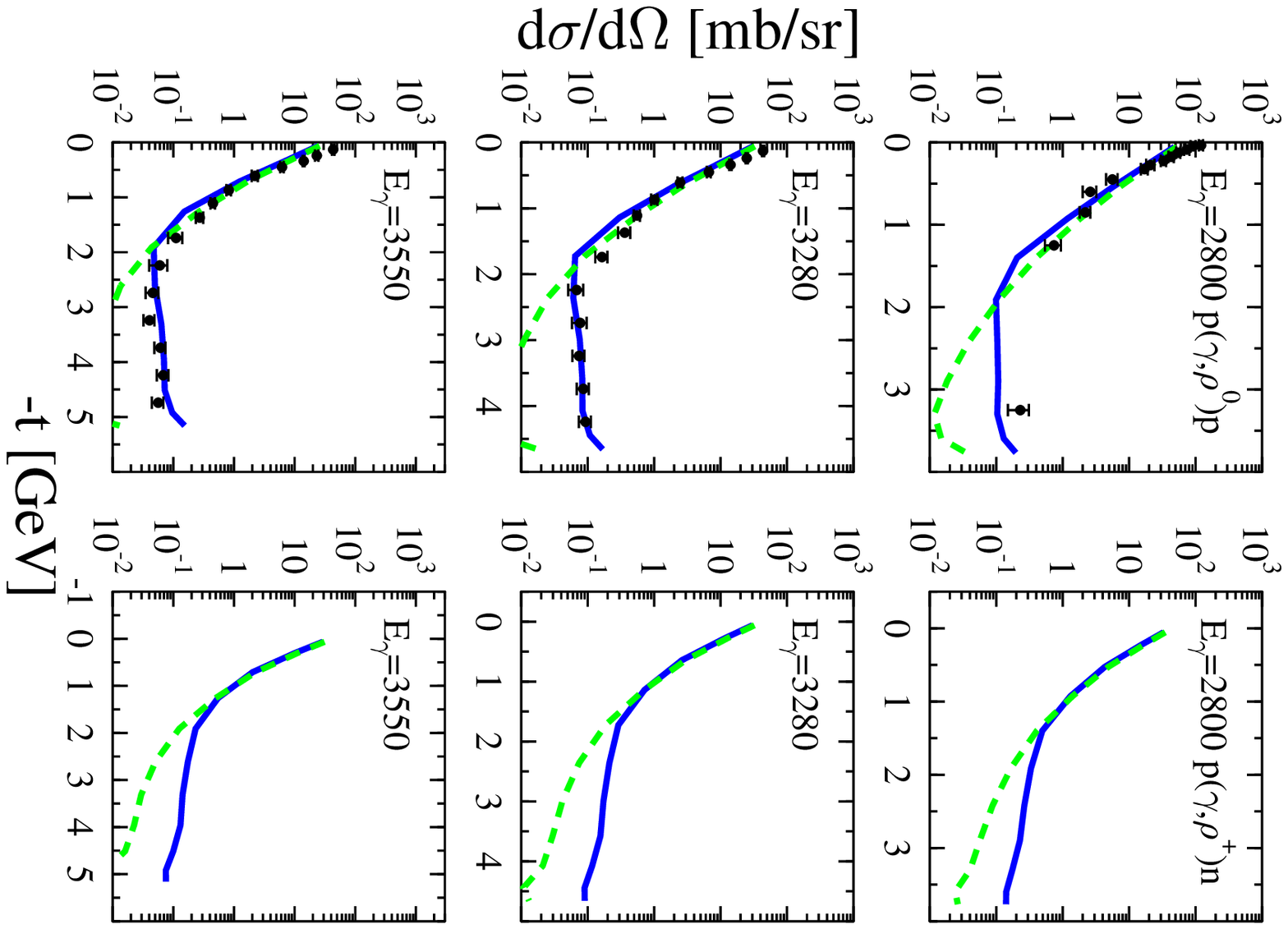}
  \caption{[Color online] Differential cross sections for
    $\rho$-meson photoproduction. The solid line corresponds to the
    full calculation. The dashed line shows the tree-level
    calculation. Experimental data are taken from ref.~\cite{Batt03}.
    The energies $E_\gamma$ are given in MeV.}
  \figlab{gam-rho-diff}
\end{figure}

\section{Results}\seclab{results}

As the dominant source of indirect contributions to the $\gamma+N \to
\rho+N$ channel we identify the $\gamma+N \to \pi+N \to \rho+N$
re-scattering process. Before addressing photo-induced $\rho$-meson
production results, we discuss the results for pion photoproduction
and pion-induced $\rho$-meson production, as these processes are
crucial for the correct description of coupled-channels effects in the
$\rho$ photoproduction channel.

In \figref{gam-pi-pw} we compare partial wave data~\cite{SAID} with
calculations for pion photoproduction at lower energies ($W \le 2
\GeV$). Shown are the results of the full calculation (darker blue
line) and the results of the calculation where the tree-level
description of the $\gamma+N \to \pi+N$ channel agrees with that of
ref.~\cite{Usov05} (lighter green line, the ``old'' calculation). Note
that due to the difference in treatment of coupled channels (primarily
$\pi+N \to \rho+N$ reaction channel) the results of the ``old''
calculation differs slightly from what was presented in
ref.~\cite{Usov05}. The present calculation, where some parameters
were refitted, gives a considerable improvement in most partial wave
amplitudes. The most important ingredients of the refitting are the
suppression of $\omega$-meson t-channel contribution and the
introduction of modified u-channel form-factors -- they have impact on
all partial waves. The softening of form-factors and introduction of
quadrupole form-factors for magnetic moment contributions are most
visible in the highest-energy tail of $E_{0+}^{3/2}$, $M_{1-}^{3/2}$,
$E_{2-}^{3/2}$ and $M_{1+}^{3/2}$ amplitudes.

The results of the refitting are illustrated in some more detail in
\figref{gam-pi-diff}.  The solid line shows the results of the final
calculation.  The calculation depicted with a dashed line uses
non-modified (dipole) u-channel form-factors, which results in the
unrealistically strong backward peaking of the differential
cross-section. The dash-dotted line in \figref{gam-pi-diff} shows the
results of the calculation in which the $g_{NN\omega}$ coupling
constant is increased to 8. This shows a too strong forward peaking.
Despite the form-factor included in the $\omega$-meson t-channel
contribution, the peak in the forward direction increases with the
energy. We have investigated the possibility of including the Regge
trajectory for the $\omega$-meson t-channel contribution, and although
it suppresses the growth of the forward peak with increasing energy,
the numerical difference in the considered energy region is small and
we decided to adopt the simpler option of using dipole-type
form-factors.  The impact of other changes, such as softening of
form-factors and the use of quadrupole form-factors for magnetic
moment contributions, is less visible in the cross-section plots, but
result in visible changes in the partial waves decomposition.

In our approach the pion-induced $\rho$ production channel is
essentially parameters-free, where only the strength of additional
gauge-invariant contact terms is adjusted. Unfortunately there exists
no direct experimental data for this channel and we rely on the
inelasticities in pion scattering in order to constrain the
parameters. As is shown in \figref{pi-pi-pw} (difference between solid
and dashed lines) the inclusion of the $\rho-N$ final state results in
the generation of substantial inelasticities in pion scattering.
The dash-dotted line in \figref{pi-pi-pw} shows the results of the
calculation where the contact term of~\eqref{pi-rho-ct3} is omitted in
the pion-induced $\rho$-meson production channel. The effect of this
contact term is to suppress the contribution of the t-channel,
especially in the higher partial waves. An unrealistically steep rise
of inelasticities in the higher partial waves shows that the inclusion
of the contact term of~\eqref{pi-rho-ct3} is vital for an accurate
description of the experimental data.

For the pion-induced $\rho$-meson production channel there exist one
more degree of ambiguity in constructing the gauge-invariance
restoring terms as compared to photo-induced pion or kaon production.
This ambiguity we express as the contact term of~\eqref{pi-rho-ct2}.
Excluding this contact term from the $\pi+N \to \rho+N$ channel we
obtain the result shown as the dotted line in~\figref{pi-pi-pw}. The
impact of this contact term can also be observed in the kaon and
$\eta$ production channels, as is shown in~\figref{gam-kl-tot},
despite the fact that these are not coupled directly. While inclusion
of this contact term seems to improve the description of $K\Lambda$
and $\eta$ production, it's effects on pion scattering and
photoproduction channels are more controversial. The largest impact is
seen in the $S_{11}$ partial wave, where the description of the
inelasticities improves at lower energies ($W \simeq 1.8 \GeV$), but
becomes worse at higher energies ($W > 2 \GeV$). The impact on the
$S_{31}$ partial wave is quite small with the current choice of
parameters, but could be of comparable strength.  In other partial
waves the effects are either small ($P_{11}$ and $P_{31}$ partial
waves) or precisely zero (spin $3/2$ and higher partial waves). In the
pion photo-production sector the effect of this contact term is
visible in $E_{0+}^{1/2}$ and $E_{0+}^{3/2}$ amplitudes at energies of
$W > 1.7 \GeV$.  In general the decision whether this contact term has
to be included should be based on the results of a global fit, which
will be presented in a forthcoming publication.

Inclusion of the contact term of~\eqref{pi-rho-ct2}, results in
increased inelasticities in the pion sector, which is reflected in a
dip in the photo-induced $K\Lambda$ production cross-section at $W =
1.7-2 \GeV$, see~\figref{gam-kl-tot}. A more careful analysis on the
level of partial waves reveals that the effect is seen primarily in S
waves. Moreover, we have found that a similar suppression effect can
be achieved by introducing an additional $S_{11}$ resonance with
appropriate parameters.  While these two mechanisms (channel coupling
effects and resonant contribution) give very similar results for the
cross-sections and final state polarization, they are clearly
distinguishable in the partial wave decomposition and could be
separated given a full set of polarization observables. Some slight
suppression of the final calculation compared to the one presented in
ref.~\cite{Usov05} (dash-dotted line) is due to somewhat increased
inelasticities in pion sector. In order to minimize differences in the
parameters we have decided to not refit kaon production channels, and
are leaving this for the forthcoming publication.

In the $\eta$ photoproduction channel, as shown in
\figref{gam-kl-tot}, the solid line represents the full calculation,
and the dash-dotted line shows the results of the ``old'' calculation.
The difference between them is primarily due to the inclusion of the
$\rho$-meson t-channel contribution. Striking feature of the current
calculation is the prominent peak at an energy of about $W = 1.74
\GeV$. To show the origin of this peak we present the results of a
calculation where the contact term of~\eqref{pi-rho-ct2} is excluded,
shown as the dashed line in~\figref{gam-kl-tot}. This clearly
demonstrates the peak is not a resonant contribution, but a channel
coupling effect. It is interesting to note that this peak coincides in
energy with the possible additional $S_{11}$ resonance, as claimed in
refs.~\cite{Sagh01,Chen03}.

In \figref{gam-rho-diff} we present the results for the $\gamma+N \to
\rho+N$ channel. The dashed line corresponds to the tree-level
calculation. Due to the use of modified u-channel form-factors and the
use of DW-style suppression scheme we are able to get a similar
description of the differential cross-section at the tree level, as
obtained in ref.~\cite{Oh04}, while using much harder form-factors.
The inclusion of Regge trajectory for the $\sigma$-meson t-channel
contribution is essential for the correct description of the forward
peak, and in the considered energy region allows to simulate the
effects of the Pomeron contribution. Including the channel coupling
effects results in the increase of the differential cross-section in
the region of high momentum transfer, as is illustrated by the solid
line. The coupled-channels result shows a good agreement with the
data.

\section{Conclusion}

We have extended the applicability of our coupled channels calculation
to energies of the order of $\sqrt{s} = 3.5 \GeV$. A good agreement
with experimental data is achieved for $\gamma+N \to \rho+N$ reaction.
We show that the channel coupling effects are extremely important in
describing the experimental data on photo-induced $\rho$-meson
production. In the kinematical regime of high momentum transfer it is
the dominant contribution. As a primary source of indirect
contributions we identify a re-scattering via an intermediate $\pi-N$
state.  Unfortunately there exist no direct data on pion-induced
$\rho$-meson production in the energy regime up to $\sqrt{s} = 3.5
\GeV$. While it is possible to use the inelasticities induced in pion
scattering and pion production channel to constrain the implementation
of pion-induced $\rho$-meson production channel, this still limits the
accuracy of the predictions, as the magnitude of the re-scattering
contributions is very sensitive to each of this channels.

The tendency of the coupled-channels calculation to enhance the
cross-section at the large angles is very important for the correct
comparison with the experimental data. As an example the comparison of
the $f_2$-meson exchange with $\sigma$-meson exchange mechanisms
presented as models ``A'' and ``B'' in ref.~\cite{Oh04}, shows that
the main differences between them lie in the high momentum transfer
region, where the coupled-channels contributions are playing dominant
role.

\end{document}